# Identifying Independent Components and Internal Process Order Parameters in Nonequilibrium Multicomponent Nonstoichiometric Compounds

Yanzhou Ji[1,2,*], Yueze Tan[2], and Long-Qing Chen[2]

[1]*Department of Materials Science and Engineering, The Ohio State University, Columbus, OH 43210, USA*

[2]*Department of Materials Science and Engineering, The Pennsylvania State University, University Park, PA 16802, USA*

**Abstract**

In CALPHAD-type thermodynamic databases, nonstoichiometric compounds are typically described by sublattice models where the sublattice site fractions represent the occupation probability of different atomic, ionic or defect species on different sublattices. Here, we develop a general procedure and corresponding linear algebra tools for converting the sublattice site fractions to a combination of independent component compositions and internal process order parameters describing the extent of internal atomic exchange, electronic redox and defect generation reactions. We apply them to a number of nonstoichiometric phases in thermodynamic databases and literature. The general procedure can be applied to constructing thermodynamic databases in terms of internal process order parameters for nonstoichiometric phases in multicomponent systems such as high-entropy oxides and alloys, which can be utilized to model their kinetics of nonequilibrium processes and microstructure evolution.

[*]Corresponding author: Yanzhou Ji. Email: ji.730@osu.edu

## 1. Introduction

Due to increased entropic contributions at finite temperatures, nonstoichiometric compounds with partial or complete disordering at different Wycoff positions or sublattices are ubiquitous. The sublattice model[1–4] and the compound energy formalism (CEF)[2,5] are often used to account for the level of nonstoichiometry at different sublattices, in which the chemical potential[6], or molar Gibbs free energy, of a nonstoichiometric compound, is expressed as a function of sublattice site fractions describing the occupation probability of different atomic, ionic or defect species on different sublattices. In fact, about 50% of the phases in commercial and literature-reported CALPHAD-based thermodynamic databases are described by sublattice models.

However, sublattice site fractions are generally not independent of each other. They must satisfy the conservation conditions for lattice sites and local charge neutrality. There are often more site fractions than the number of independent components[3]. Furthermore, obtaining the chemical potentials of components in nonstoichiometric phases generally requires complicated iterative numerical solvers[7]. However, as we have shown in our recent study[8], the extra degrees of freedom within the sublattice model of a nonstoichiometric phase may correspond to different types of internal processes, which can be described by a set of internal process order parameters (IPOPs) representing the extent of these internal processes. Therefore, the sublattice site fractions can be fully converted to a combination of independent component compositions and IPOPs, and the converted chemical potentials, now as a function of compositions and IPOPs, can be directly used in thermodynamic analysis and kinetic modeling. We have previously discussed the theoretical basis, general requirement and procedure to identify the internal processes and the IPOPs for the conversion[8]. However, a general multicomponent nonstoichiometric phase with multiple sublattices may involve multiple independent internal processes.

In this study, we extend the applicability range of our proposed thermodynamic description of nonstoichiometric phases using IPOPs and develop the numerical tools to facilitate the conversion. These tools can then be applied to various nonstoichiometric phases in existing CALPHAD-type thermodynamic databases and literature to establish thermodynamic databases in terms of IPOPs to model their nonequilibrium processes and microstructure evolution in multicomponent systems such as high-entropy alloys and oxides.

## 2. Methods

We first introduce the theoretical basis for converting sublattice site fractions to independent component compositions and IPOPs, and then present the general procedure, numerical tools and its implementation.

### *2.1 Theoretical basis*

For a nonstoichiometric compound described by a sublattice model, the Gibbs free energy is typically formulated in terms of sublattice site fractions. The differential form of the fundamental equation of thermodynamics in terms of the Gibbs free energy is given by[9]

$$dG = -SdT + Vdp + \sum_{i,j,\cdots,m} \mu_{i:j:\cdots:m} dN_{i:j:\cdots:m} = -SdT + Vdp + \sum_{j=1}^{n_s}\sum_{i=1}^{m_j} \mu_i^{(j)} dN_i^{(j)} \quad (1a)$$

where $T$ is temperature, $S$ is entropy, $p$ is pressure, $V$ is volume; $i:j:\cdots:m$ indicates an endmember with ":" separating the different sublattices, $\mu_{i:j:\cdots:m}$ is the chemical potential of the endmember and $N_{i:j:\cdots:m}$ is the number of moles of the endmember; $\mu_i^{(j)}$ and $N_i^{(j)}$ are the chemical potential and the number of moles of the $i$th species in the $j$th sublattice; $n_s$ is the total number of sublattices and $m_j$ is the total number of species in the $j$th sublattice.

However, site fractions are usually not pure external thermodynamic variables like component compositions. They can be related to component compositions, but may involve extra degrees of freedom depending on the specific form of the sublattice model. For example, for a nonstoichiometric phase $(A,B)_1(B)_1$, the site fractions $y_A^{(1)}$, $y_B^{(1)}$, and $y_B^{(2)}$ can be readily converted to the component composition $x_B$ by solving $y_A^{(1)} + y_B^{(1)} = 1$, $y_B^{(2)} = 1$, and $x_B = \frac{y_B^{(1)} + y_B^{(2)}}{2}$. However, for $(A,B)_1(A,B)_1$, there is one extra degree of freedom in site fractions since we can only list 3 constraints ($y_A^{(1)} + y_B^{(1)} = 1$, $y_A^{(2)} + y_B^{(2)} = 1$, and $x_B = \frac{y_B^{(1)} + y_B^{(2)}}{2}$) while there are 4 site fractions. Indeed, site fractions are of mixed external and internal nature. Therefore, for a nonstoichiometric phase with extra degrees of freedom in its sublattice model, we can convert Eq. (1a) into

$$dG = -SdT + Vdp + \sum_i \mu_i dN_i - \sum_j D_j d\xi_j \quad (1b)$$

where $\mu_i$ and $N_i$ are the chemical potential and amount of component $i$, respectively, $\xi_j$ is an internal variable or IPOP describing the extent of the $j$th internal process, and $D_j$ is its driving force given by

$$D_j = -\left(\frac{\partial G}{\partial \xi_j}\right)_{T,p,N_i,\xi_{k\neq j}} \quad (2)$$

The number of independent internal processes or IPOPs depends on the number of site fractions and their external constraints including site conservation, charge neutrality and their relations to component compositions, which we will discuss in detail in later sections.

It should be noted that for such nonstoichiometric phases, given temperature, pressure and component compositions, a commonly used approach to calculate the site fractions and chemical potential $\mu_i$ is to assume internal equilibrium, i.e., $D_j = 0$ for all $j$s. This approach has been widely implemented in CALPHAD-based software packages such as Thermo-Calc, Pandat, OpenCalphad[10] and pycalphad[11]. The internal equilibrium assumption would be valid if the internal processes are much faster than solute diffusion and phase transformation. However, kinetically, it is also of interest to explore how these internal processes affect the phase stabilities and phase transformation pathways, where $D_j = 0$ (for all $j$s) no longer holds. Therefore, under general nonequilibrium conditions, the $-\sum_j D_j d\xi_j$ term would not vanish in Eq. (1b) and $\sum_i \mu_i dN_i$ alone would not be sufficient to fully represent the $\sum_{i,j,\cdots,m} \mu_{i:j:\cdots:m} dN_{i:j:\cdots:m}$ or $\sum_{j=1}^{n_s} \sum_{i=1}^{m_j} \mu_i^{(j)} dN_i^{(j)}$ terms in Eq. (1a). Meanwhile, to calculate site fractions under general nonequilibrium conditions, component compositions alone are insufficient; we also need the values of all $\xi_j$s. Therefore, it is critical to physically identify the internal processes and properly define their order parameters ($\xi_j$s) to achieve the correct internal equilibrium.

The internal processes within a nonstoichiometric phase may take place among different sublattice species. Here we would like to further emphasize the difference between "components" and "species". In thermodynamics, a component is a chemically independent constituent of a system while a species that participates in an internal process may not be chemically independent. For example, a chemical element A can be considered as a component, but the component A may manifest itself with different chemical species such as atoms (A), ions ($A^+$, $A^{2+}$, etc.), or molecules ($A_2$, $A_3$, etc.), while these different chemical species are not independent of each other. Note that vacancy (Va) is not a component. Internal processes in a nonstoichiometric phase can generally be classified into different types and expressed in the form of reactions[8]: (i) internal atomic exchange reactions; (ii) internal redox reactions; (iii) internal defect generation reactions; (iv) internal molecule formation reaction. Each internal reaction can be expressed as a reaction among different species on the same or different sublattices:

$$\sum_{i,j} v_{R_i^{(j)}} R_i^{(j)} = \sum_{m,n} v_{P_m^{(n)}} P_m^{(n)} \tag{3}$$

where $R_i^{(j)}$ and $P_m^{(n)}$ are reacting and product species, $v_{R_i^{(j)}}$ and $v_{P_m^{(n)}}$ are their stoichiometries, with the subscripts (*i* and *m*) denoting the indices of species and the superscripts ((*j*) and (*n*)) denoting the indices of sublattices. To clearly illustrate the different types of internal processes, we consider a hypothetical B2-ordered nonstoichiometric phase with two components A and B, which have two sublattices: one at the corners and the other at the body centers of the cubic crystal.

(i) If the nonstoichiometric phase is described by a sublattice model (A, B)₁(A, B)₁, then as illustrated in Figure 1, there is an atom exchange reaction between the A and B atoms in the two sublattices and Eq. (3) can be written as

$$A^{(1)} + B^{(2)} = A^{(2)} + B^{(1)} \tag{4i}$$

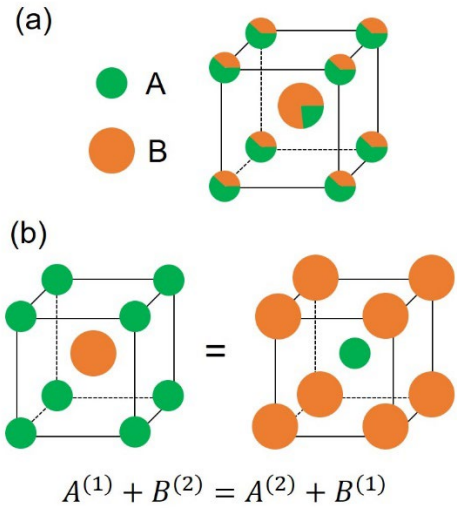

Figure 1: Illustration of a simple B2-ordered nonstoichiometric phase (A, B)₁(A, B)₁ and its internal atom exchange reaction. (a) The crystal structure and illustration of sublattice species distributions. (b) Illustration of the atom exchange reaction.

(ii) If the nonstoichiometric phase is described by a sublattice model $(A^+, A^{2+}, A^{3+})_1(B^{2-}, Va)_1$, then as illustrated in Figure 2, there is a redox reaction among the three cations of the component A on the first sublattice and Eq. (3) can be written as

$$(A^+)^{(1)} + (A^{3+})^{(1)} = 2(A^{2+})^{(1)} \tag{4ii}$$

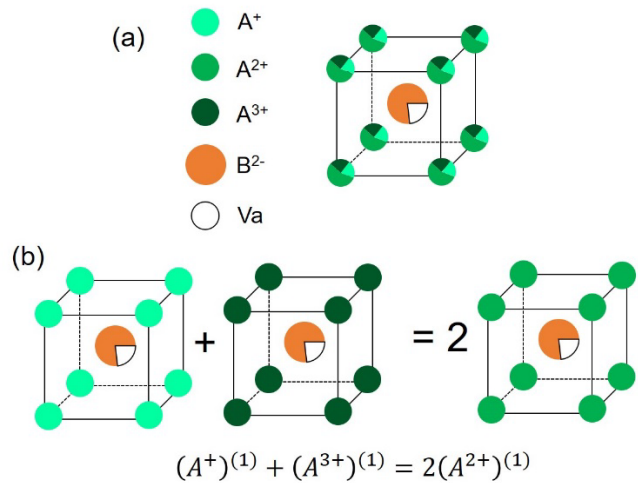

$(A^+)^{(1)} + (A^{3+})^{(1)} = 2(A^{2+})^{(1)}$

Figure 2: Illustration of a simple B2-ordered nonstoichiometric phase $(A^+, A^{2+}, A^{3+})_1(B^{2-}, Va)_1$ and its internal redox reaction. (a) The crystal structure and illustration of sublattice species distributions. (b) Illustration of the redox reaction.

(iii)    If the nonstoichiometric phase is described by a sublattice model $(A, Va)_1(B, Va)_1$, then as illustrated in Figure 3, a Schottky defect can be generated by the vacancy sites on the two sublattices, and Eq. (3) can be written as

$$Null = Va^{(1)} + Va^{(2)} \qquad (4\text{iii})$$

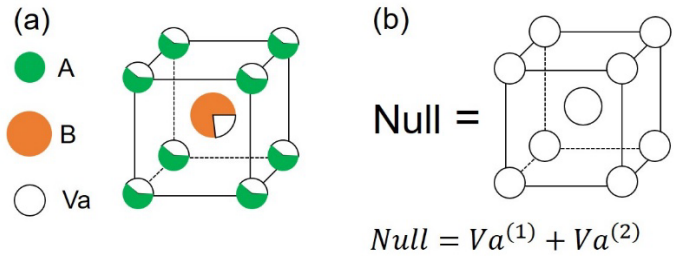

Figure 3: Illustration of a simple B2-ordered nonstoichiometric phase $(A, Va)_1(B, Va)_1$ and its internal defect generation reaction. (a) The crystal structure and illustration of sublattice species distributions. (b) Illustration of the defect generation reaction to form a Schottky defect.

Alternatively, if the nonstoichiometric phase is described by a sublattice model $(A, B)_1(A, Va)_1$, then as illustrated in Figure 4, the A atom on the first sublattice can migrate to the initially vacant second sublattice, which is analogous to the formation of the interstitial site of a Frenkel defect. Eq. (3) can be written as

$$A^{(1)} + 2Va^{(2)} = A^{(2)} \qquad (4\text{iv})$$

Note that this internal reaction describes the Frenkel defect formation if the second sublattice is an interstitial site. In general, this type of reaction can also take place if both lattice sites are substitutional sites.

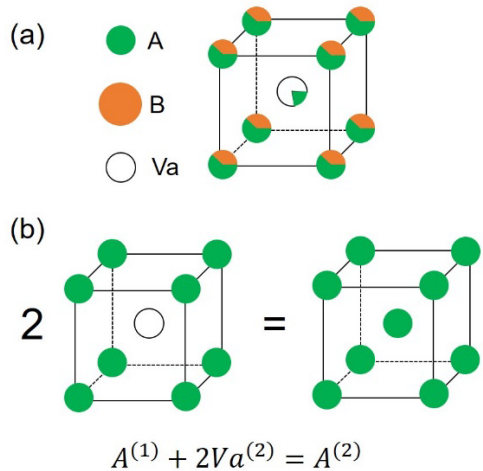

$$A^{(1)} + 2Va^{(2)} = A^{(2)}$$

Figure 4: Illustration of a simple B2-ordered nonstoichiometric phase $(A, B)_1(A, Va)_1$ and its internal defect generation reaction. (a) The crystal structure and illustration of sublattice species distributions. (b) Illustration of the defect generation reaction to form a Frenkel-like defect.

(iv) If the nonstoichiometric phase is described by a sublattice model $(A, B)_1(A, AB)_1$, then as illustrated in Figure 5, the molecule species "AB" consisting of both A and B atoms in the second sublattice can be formed via atomic species "A" or "B". Eq. (3) can be written as

$$2B^{(1)} + 3A^{(2)} = A^{(1)} + 2AB^{(2)} \qquad (4v)$$

It should be noted that the molecule species in a sublattice are most commonly considered in gas or liquid phases. However, in some nonstoichiometric solid phases, molecular species are also reported. For example, in the nonstoichiometric $Co_{17}Y_2$ phase[12], a $Co_2$ dimer is considered to be able to replace a Y atom in the first two sublattices, and the sublattice model is formulated as $(Co_2, Y)_1(Co_2, Y)_2(Co)_{15}$.

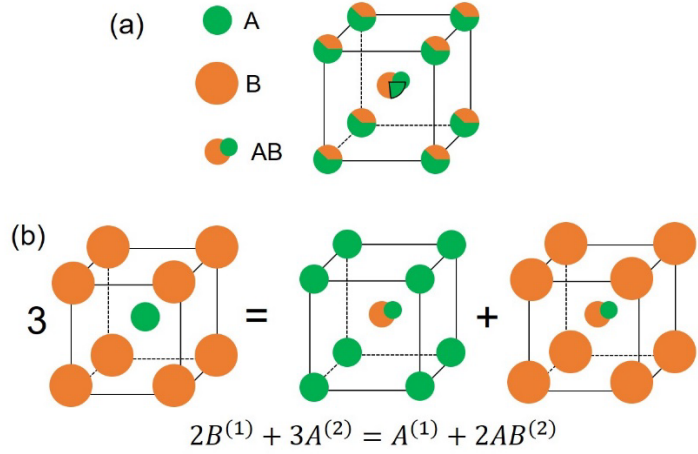

$$2B^{(1)} + 3A^{(2)} = A^{(1)} + 2AB^{(2)}$$

Figure 5: Illustration of a simple B2-ordered nonstoichiometric phase $(A, B)_1(A, AB)_1$ and its internal molecule formation reaction. (a) The crystal structure and illustration of sublattice species distributions. (b) Illustration of the molecule formation reaction.

The driving force $D'$ for the internal reaction is

$$D' = \sum_{m,n} \nu_{P_m^{(n)}} \mu_{P_m^{(n)}} - \sum_{i,j} \nu_{R_i^{(j)}} \mu_{R_i^{(j)}} \qquad (5)$$

where $\mu_{R_i^{(j)}}$ and $\mu_{P_m^{(n)}}$ are the chemical potentials of $R_i^{(j)}$ and $P_m^{(n)}$, respectively.

With the internal processes identified, one need to properly define the form of $\xi$ so that $D \propto D'$, which ensures $D = D' = 0$ at equilibrium. Moreover, to represent the extent of the internal process, $\xi$ should range from 0 (full reactant and no product) to 1 (full product and no reactant)[13]. In addition, the form of $\xi$ should not artificially alter the phase stability. We have shown that[8], to satisfy all these requirements, $\xi$ should be a function of the site fractions of the species that participate in the internal reaction, i.e., $\xi = \xi\left(\left\{y_{R_i}^{(j)}\right\}, \left\{y_{P_m}^{(n)}\right\}\right)$, where we use $y$ to represent the site fraction with its subscript and superscript representing the index of the species and the index of the sublattice, respectively. The simplest and most straightforward expression of IPOPs would be the site fractions themselves, or their linear combinations, which is sufficient and convenient for most situations. Alternatively, to more clearly indicate the internal process, we would suggest to use[8]

$$\xi = \frac{\sum_{m,n} y_{P_m}^{(n)}}{\sum_{i,j} y_{R_i}^{(j)} + \sum_{m,n} y_{P_m}^{(n)}} \quad (6)$$

It should be noted that the reactant of a defect generation reaction can be "null". For example, a physically meaningful $\xi$ expression for vacancy (Va) generation should be

$$\xi = \sum_{j} v_{Va}^{(j)} y_{Va}^{(j)} \quad (7)$$

As such, the specific form of $\xi$ can be determined in terms of site fractions. We emphasize that although $\xi$ depends on site fractions, it is independent of component compositions; the combination of independent component compositions and IPOPs can fully represent the site fractions.

## 2.2 *Conversion procedure and numerical tools*

Let us consider a nonstoichiometric phase with $n$ components, $s$ sublattices and $m_i$ ($i=1,2,…,s$) species on each sublattice. The stoichiometry of each sublattice is $k_i$ ($i=1,2,…,s$). To convert $\{y_i^{(j)}\}$ to a combination of independent component compositions $\{x_l\}$ and IPOPs $\{\xi_m\}$, we propose the following procedure:

(1) List all the external constraints for the sublattice site fractions. These include $s$ lattice site conservation conditions,

$$\sum_{i=1}^{m_j} y_i^{(j)} = 1, (j = 1,2,\cdots,s) \quad (8)$$

one (or zero, if all the species in the sublattice model are charge-neutral) charge neutrality condition,

$$\sum_{j=1}^{s}\sum_{i=1}^{m_j} k_j z_i^{(j)} y_i^{(j)} = 0 \tag{9}$$

where $z_i^{(j)}$ is the valence of the $i$-th species in the $j$-th sublattice, and $n$ relations between component compositions, and site fractions,

$$x_l = \frac{\sum_{j=1}^{s}\sum_{i=1, i\in l}^{m_j} k_j y_i^{(j)}}{M} \quad (l = 1, 2, \cdots, n) \tag{10}$$

where $M = \sum_{j=1}^{s} k_j \sum_{i=1, i\notin\{Va, e', h^{\cdot}\}}^{m_j} y_i^{(j)}$ is the total number of moles in a formula unit. Here "$i \notin \{Va, e', h^{\cdot}\}$" means that the vacancies, electrons ($e'$) and holes ($h^{\cdot}$) are not considered as components and are not counted when calculating $M$. Note that these $n$ relations may not be fully independent of each other because they are at least constrained by a mass conservation condition $\sum_{l=1}^{n} x_l = 1$. The number of independent components is therefore typically $n$-1 due to this mass conservation constraint, but this number can be even less depending on the specific sublattice model. These conditions will give $c$ independent constraints, with $c<s+n+1$.

(2) To identify the internal constraints, list all possible internal reactions within the system (e.g., Eq. (3)), and write all the corresponding $\xi$ expressions according to Eqs. (6-7). Note the total number of independent reactions should be $\sum_{i=1}^{s} m_i - c$, while the number of internal reactions listed here, $m'$, may be larger than this value. To exhaustively list the internal reactions within the system, we follow the typical "patterns" of the different types of internal processes in the sublattice model, and find all the species pairs and reactions that match the patterns. To be more specific,

(i) The typical "pattern" of the internal atom exchange reaction in the sublattice model of a nonstoichiometric phase is that there exist at least one pair of species that are present in different sublattices. Taking the Figure 1 example, the species pair is (A, B), which is present in both the first and the second sublattices. More generally, the species pair can belong to either the same or different components; one of the species in the species pair can be vacancy. A multicomponent nonstoichiometric phase can have multiple such species pairs in more than two sublattices.

(ii) For the internal redox reaction, the typical pattern is that there are at least two different pairs of species of the same component with different valences, in the same or different sublattices. For example, in Figure 2, the species pairs are $(A^+, A^{2+})$ and $(A^{2+}, A^{3+})$, both in the first sublattice. More generally, the two species pairs can come from different components in different sublattices, e.g., the first pair is $(A^+, A^{2+})$ in the first sublattice and the second pair is $(B^+, B^{2+})$ in the second sublattice.

(iii) The pattern of the Schottky defect generation reaction is that vacancy sites are present in all sublattices to form a fully vacant lattice (i.e., the Schottky defect), as shown in Figure 3. The pattern of the Frenkel-like defect generation reaction is that given a species in one sublattice (e.g., the species A in the first sublattice in Figure 4), it is also present in another sublattice along with vacancy (e.g., the second sublattice in Figure 4 contains both species A and vacancy).

(iv) The pattern of the molecule formation reaction is that there are both atomic or molecular species within the system. For example, in Figure 5, the molecule species "AB" in the second sublattice will form from the internal reaction.

The typical patterns of these internal processes are summarized in Table 1.

Table 1: "Patterns" of internal processes in the sublattice model

| Internal processes | "Patterns" |
| --- | --- |
| Atom exchange reaction | The same pair of species in different sublattices. |
| Redox reaction | Different pairs of species of the same component with different valences, in the same or different sublattices. |
| Defect generation reaction | Vacancy sites are present in all sublattices; or given a species in one sublattice, it is also present in another sublattice along with vacancy. |
| Molecule formation reaction | Molecule species are present in the system. |

For a nonstoichiometric phase with a given sublattice model, we can exhaustively list all the reactions that follow the patterns in Table 1. If a nonstoichiometric phase includes more than one internal processes, we can write these internal reactions into matrix form **Au=0**, where **A** is a $m' \times \sum_{i=1}^{s} m_i$ coefficient matrix, and **u** is a vector of species consisting of $\sum_{i=1}^{s} m_i$ rows[8]. For example, consider a nonstoichiometric phase described by the sublattice model (A, B, C)$_1$(A, B, C)$_1$ with $s=2$ sublattices and $\sum_{i=1}^{s} m_i = 6$ sublattice species. Following the patterns above, this system has three species pairs (A, B), (A, C) and (B, C) in both sublattices, indicating $m'=3$ possible atom exchange reactions $A^{(1)} + B^{(2)} = B^{(1)} + A^{(2)}$, $A^{(1)} + C^{(2)} = C^{(1)} + A^{(2)}$, and $B^{(1)} + C^{(2)} = C^{(1)} + B^{(2)}$. These three internal reactions can be written as

$$\begin{pmatrix} 1 & -1 & 0 & -1 & 1 & 0 \\ 1 & 0 & -1 & -1 & 0 & 1 \\ 0 & 1 & -1 & 0 & -1 & 1 \end{pmatrix} \begin{pmatrix} A^{(1)} \\ B^{(1)} \\ C^{(1)} \\ A^{(2)} \\ B^{(2)} \\ C^{(2)} \end{pmatrix} = \mathbf{Au} = 0 \qquad (11)$$

where the positive elements in **A** indicate the reactant species while the negative elements indicate product species.

The independent combinations of the internal reactions can be identified by finding the submatrices of **A** with $\sum_{i=1}^{s} m_i - c$ rows and $\sum_{i=1}^{s} m_i$ columns and the rank of $\sum_{i=1}^{s} m_i - c$. Note that the desirable submatrices of **A** may not be unique, but equivalent in the sense that they represent the independent combinations of internal reactions. For example, in Eq. (11), the rank of **A** is 2, indicating two independent internal reactions. The desirable submatrices of **A** with two rows can be obtained by selecting any two of the three rows of **A**. Alternatively, we can just leave the total *m'* internal reactions and internal process order parameter expressions in this step and identify their independent combinations in the next step.

(3)    With all the external and internal constraints listed, we can further write these constraint conditions, including Eqs. (6-7) and Eqs. (8-10) into matrix form, i.e., **Bv=f**, where **v** is a vector of site fractions of sublattice species consisting of $\sum_{i=1}^{s} m_i$ rows, **B** is a known coefficient matrix of dimension $(s + n + 1 + \sum_{i=1}^{s} m_i - c) \times \sum_{i=1}^{s} m_i$ (or $(s + n + 1 + m') \times \sum_{i=1}^{s} m_i$ if the independent internal reactions have not been identified yet) and **f** is a known vector. In **v**, we follow the definition order of the sublattice site fractions. For example, for $(A, B, C)_1(A, B, C)_1$, we define $\mathbf{v} = \left(y_A^{(1)}, y_B^{(1)}, y_C^{(1)}, y_A^{(2)}, y_B^{(2)}, y_C^{(2)}\right)^T$.

In the first *s* rows of the coefficient matrix **B** and the vector **f**, we put the *s* site conservation conditions (Eq. (8)), i.e., the $i^{\text{th}}$ ($i \leq s$) row of **B** corresponds to the site conservation condition for the $i^{\text{th}}$ sublattice. Therefore, following the order of the site fractions in **v**, the $j^{\text{th}}$ element in the $i^{\text{th}}$ row of B is either 1 (if $\sum_{l=1}^{i-1} m_l < j \leq \sum_{l=1}^{i} m_l$) or 0 (if $j > \sum_{l=1}^{i} m_l$ or $j \leq \sum_{l=1}^{i-1} m_l$). Correspondingly, the first $s$ elements of the vector $\mathbf{f}$ are all 1 since the right-hand side of Eq. (8) is 1. For example, for (A, B, C)$_1$(A, B, C)$_1$, since the site conservation conditions are $y_A^{(1)} + y_B^{(1)} + y_C^{(1)} = 1$ and $y_A^{(2)} + y_B^{(2)} + y_C^{(2)} = 1$, the first two rows of the $\mathbf{B}$ matrix can be expressed as

$$\begin{pmatrix} 1 & 1 & 1 & 0 & 0 & 0 \\ 0 & 0 & 0 & 1 & 1 & 1 \end{pmatrix}$$

while the first two rows of the $\mathbf{f}$ vector can be written as $(1,1)^T$.

In the $(s+1)^{\text{th}}$ row of $\mathbf{B}$ and $\mathbf{f}$ we put the charge neutrality condition (Eq. (9)). Therefore, each element of $\mathbf{B}$ in this row is the valence of the sublattice species $z_q^p$ multiplied by the stoichiometry of the sublattice $k_p$, if this element corresponds to the $q^{\text{th}}$ species in the $p^{\text{th}}$ sublattice. The $(s+1)^{\text{th}}$ row of $\mathbf{f}$ is 0 due to charge neutrality. If the nonstoichiometric phase only involves charge-neutral species such as (A, B, C)$_1$(A, B, C)$_1$, then this $(s+1)^{\text{th}}$ row is neglected.

In the next $n$ rows, we put the $n$ definition equations of component compositions. Here we re-arrange Eq. (10) as

$$\sum_{j=1}^{s} \sum_{i=1, i \in l}^{m_j} k_j y_i^{(j)} - Mx_l = \sum_{j=1}^{s} k_j \left( \sum_{i=1, i \in l}^{m_j} y_i^{(j)} - x_l \sum_{i=1, i \notin \{Va, e', h\}}^{m_j} y_i^{(j)} \right) = 0 \quad (12)$$

$$(l = 1, 2, \cdots, n)$$

so that the left-hand side of these relations only include linear terms of site fractions. All these $n$ rows of $\mathbf{f}$ are 0 since the right-hand side of Eq. (12) is 0. According to Eq. (12), in the $l^{\text{th}}$ row of these $n$ rows of the matrix $\mathbf{B}$, if a species in the $p^{\text{th}}$ sublattice belongs to the $l^{\text{th}}$ component, then its corresponding element in this row of $\mathbf{B}$ should be $k_p(1 - x_l)$; otherwise, if the species is either vacancy, electron or hole, then the corresponding $\mathbf{B}$ element should be 0, because these species are not counted as components; finally, if this species does not belong to the $l^{th}$ component but belongs to another component, then the corresponding B element should be $-x_l k_p$. For example, for $(A, B, C)_1(A, B, C)_1$, we can define the composition of two independent components A and B as

$$x_A = \frac{y_A^{(1)} + y_A^{(2)}}{y_A^{(1)} + y_B^{(1)} + y_C^{(1)} + y_A^{(2)} + y_B^{(2)} + y_C^{(2)}} \tag{13a}$$

$$x_B = \frac{y_B^{(1)} + y_B^{(2)}}{y_A^{(1)} + y_B^{(1)} + y_C^{(1)} + y_A^{(2)} + y_B^{(2)} + y_C^{(2)}} \tag{13b}$$

which can be re-arranged following Eq. (12) as

$$(1 - x_A)y_A^{(1)} - x_A y_B^{(1)} - x_A y_C^{(1)} + (1 - x_A)y_A^{(2)} - x_A y_B^{(2)} - x_A y_C^{(2)} = 0 \tag{14a}$$
$$-x_B y_A^{(1)} + (1 - x_B)y_B^{(1)} - x_B y_C^{(1)} - x_B y_A^{(2)} + (1 - x_B)y_B^{(2)} - x_B y_C^{(2)} = 0 \tag{14b}$$

Therefore, the corresponding rows of **f** should be $(0,0)^T$ and the corresponding rows of **B** are:

$$\begin{pmatrix} 1 - x_A & -x_A & -x_A & 1 - x_A & -x_A & -x_A \\ -x_B & 1 - x_B & -x_B & -x_B & 1 - x_B & -x_B \end{pmatrix}$$

Finally, in the last $m'$ rows, we put the $m'$ definition equations for the IPOPs according to Eqs. (6-7). If the $i^{th}$ internal process is a defect generation reaction, then according to the right-hand side of Eq. (7), the corresponding row of **f** should be $\xi_i$; in the corresponding row of **B**, if a species in the $p^{th}$ sublattice is vacancy, then its **B** element is $v_{Va}^{(p)} = \frac{k_p}{\sum_{l=1}^{s} k_l}$, otherwise the **B** element is 0. If the internal process is not a defect process, then we re-write Eq. (6) as

$$(1 - \xi) \sum_{m,n} y_{P_m}^{(n)} - \xi \sum_{i,j} y_{R_i}^{(j)} = 0 \tag{15}$$

Therefore, the corresponding row of **f** should be 0. For **B**, if a species is a product of the internal process, then its B element is $(1 − \xi)$; if a species is a reactant of the internal process, then its B element is $−\xi$; if the species does not participate in the internal process, then its B element is 0. For example, for (A, B, C)₁(A, B, C)₁, from the rank of matrix **A** in Eq. (11), we can identify two independent internal processes, e.g., atom exchange reactions $A^{(1)} + B^{(2)} = B^{(1)} + A^{(2)}$ and $A^{(1)} + C^{(2)} = C^{(1)} + A^{(2)}$. The IPOPs can be defined as

$$\xi_1 = \frac{y_B^{(1)} + y_A^{(2)}}{y_A^{(1)} + y_B^{(2)} + y_A^{(2)} + y_B^{(1)}} \tag{16a}$$

$$\xi_2 = \frac{y_C^{(1)} + y_A^{(2)}}{y_A^{(1)} + y_C^{(2)} + y_A^{(2)} + y_C^{(1)}} \tag{16b}$$

which can be re-arranged according to Eq. (15) as

$$-\xi_1 y_A^{(1)} + (1 - \xi_1)y_B^{(1)} + (1 - \xi_1)y_A^{(2)} - \xi_1 y_B^{(2)} = 0 \tag{17a}$$

$$-\xi_2 y_A^{(1)} + (1 - \xi_2)y_C^{(1)} + (1 - \xi_2)y_C^{(2)} - \xi_2 y_B^{(2)} = 0 \tag{17b}$$

Therefore, the last two rows of **f** should be $(0,0)^T$ and the last two rows of **B** are:

$$\begin{pmatrix} -\xi_1 & 1-\xi_1 & 0 & 1-\xi_1 & -\xi_1 & 0 \\ -\xi_2 & 0 & 1-\xi_2 & 1-\xi_2 & 0 & -\xi_2 \end{pmatrix}$$

Taking all the conditions above, the **B** matrix should be expressed as

$$B_{ij} = \begin{cases} 1; \ 1 \leq i \leq s, \sum_{l=1}^{i-1} m_l < j \leq \sum_{l=1}^{i} m_l \\ z_q^p k_p; \ i = s+1, j = q + \sum_{l=1}^{p-1} m_l, 1 \leq q \leq m_p \\ (1-x_{i-s-1})k_p; \ s+1 < i \leq s+1+n, j = q + \sum_{l=1}^{p-1} m_l, 1 \leq q \leq m_p, q \in i-s-1 \\ -x_{i-s-1}k_p; \ s+1 < i \leq s+1+n, j = q + \sum_{l=1}^{p-1} m_l, 1 \leq q \leq m_p, q \notin i-s-1, q \notin \{Va, e', h\} \\ 1 - \xi_{i-s-1-n}; \ i > s+1+n, 1 \leq j \leq \sum_{l=1}^{s} m_l, u_j \in \{P_p^q\}^{i-s-1-n}, \{P_p^q\}^{i-s-1-n} \neq \{Va^q\}^{i-s-1-n} \\ -\xi_{i-s-1-n}; \ i > s+1+n, 1 \leq j \leq \sum_{l=1}^{s} m_l, u_j \in \{R_p^q\}^{i-s-1-n}, \{R_p^q\}^{i-s-1-n} \neq null \\ \frac{k_p}{\sum_{l=1}^{s} k_l}; \ i > s+1+n, 1 \leq j \leq \sum_{l=1}^{s} m_l, u_j \in \{P_p^q\}^{i-s-1-n} = \{Va^q\}^{i-s-1-n}, \{R_p^q\}^{i-s-1-n} = null \\ 0; else \end{cases} \quad (18)$$

And the vector **f** should be

$$f_i = \begin{cases} 1; \ 1 \leq i \leq s \\ \xi_{i-s-1-n}; \ i > s+1+n, 1 \leq j \leq \sum_{l=1}^{s} m_l, u_j \in \{P_p^q\}^{i-s-1-n} = \{Va^q\}^{i-s-1-n}, \{R_p^q\}^{i-s-1-n} = null \\ 0; \ else \end{cases} \quad (19)$$

The coefficient matrix **B** now may still contain redundant rows. To identify the $\sum_{i=1}^{S} m_i$ independent constraints, one needs to find all submatrices of **B** with $\sum_{i=1}^{S} m_i$ columns and $\sum_{i=1}^{S} m_i$ rows, and with a rank of $\sum_{i=1}^{S} m_i$. The sublattice site fractions can then be fully converted to independent component compositions and internal process order parameters by analytically solving $\mathbf{v} = \mathbf{B}^{-1}\mathbf{f}$, which gives the analytical expressions of site fractions as a function of component compositions and internal process order parameters, i.e., $y_i^{(j)} = y_i^{(j)}(\{x_l\}, \{\xi_m\})$. For example, for $(A, B, C)_1(A, B, C)_1$, the full form of the matrix **B** is

$$\mathbf{B} = \begin{pmatrix} 1 & 1 & 1 & 0 & 0 & 0 \\ 0 & 0 & 0 & 1 & 1 & 1 \\ 1-x_A & -x_A & -x_A & 1-x_A & -x_A & -x_A \\ -x_B & 1-x_B & -x_B & -x_B & 1-x_B & -x_B \\ -\xi_1 & 1-\xi_1 & 0 & 1-\xi_1 & -\xi_1 & 0 \\ -\xi_2 & 0 & 1-\xi_2 & 1-\xi_2 & 0 & -\xi_2 \end{pmatrix} \quad (20)$$

and the vector **f** is $\mathbf{f} = (1,1,0,0,0,0)^T$. Therefore, we can obtain

$$\mathbf{v} = \begin{pmatrix} y_A^{(1)} \\ y_B^{(1)} \\ y_C^{(1)} \\ y_A^{(2)} \\ y_B^{(2)} \\ y_C^{(2)} \end{pmatrix} = \mathbf{B}^{-1}\mathbf{f} = \begin{pmatrix} \frac{1}{3}(1 - 2(\xi_1 - 2)x_A + 2\xi_2(x_B - 1) - 2\xi_1 x_B) \\ \frac{1}{3}(1 + 2(2\xi_1 - 1)x_A + 2\xi_2(x_B - 1) + 4\xi_1 x_B) \\ \frac{1}{3}(1 - 2(\xi_1 + 1)x_A - 4\xi_2(x_B - 1) - 2\xi_1 x_B) \\ \frac{1}{3}(-1 + 2(\xi_1 + 1)x_A - 2\xi_2(x_B - 1) + 2\xi_1 x_B) \\ \frac{1}{3}(5 - 2(2\xi_1 - 1)x_A - 2\xi_2(x_B - 1) - 4\xi_1 x_B) \\ \frac{1}{3}(5 + 2(\xi_1 - 2)x_A + 4\xi_2(x_B - 1) + 2(\xi_1 - 3)x_B) \end{pmatrix} \quad (21)$$

With this conversion, one can further analytically obtain the chemical potential of the nonstoichiometric phase $\mu(\{x_l\}, \{\xi_m\})$ from $\mu\left(\{y_i^{(j)}\}\right)$, as well as its first derivatives and second derivatives with respect to compositions and order parameters, respectively, for further thermodynamic and kinetic analysis[8]. Note that the key to obtain the chemical potential expressions is still to analytically reconstruct $\mu\left(\{y_i^{(j)}\}\right)$ from the CALPHAD thermodynamic database of the system with detailed analytical expressions of the lattice stabilities, the standard chemical potentials of the endmembers, and the interaction parameters, while $\mu(\{x_l\}, \{\xi_m\})$ is obtained from $\mu\left(\{y_i^{(j)}\}\right)$ simply via variable conversion. We have shown such a conversion example in Eqs. (25-26) of Ref.[8].

In practice, however, the analytical expression of $y_i^{(j)} = y_i^{(j)}(\{x_l\}, \{\xi_m\})$ may be quite complicated, which may lead to a quite long analytical expression of $\mu(\{x_l\}, \{\xi_m\})$ and difficulties in analytically expressing its derivatives. To avoid such an issue, one may want to calculate these quantities numerically without knowing the detailed analytical expressions. To do so, one may start from writing the derivatives using the chain rule, i.e.,

$$\frac{\partial \mu}{\partial x_l} = \sum_{j=1}^{s} \sum_{i=1}^{m_j} \frac{\partial \mu}{\partial y_i^{(j)}} \frac{\partial y_i^{(j)}}{\partial x_l} \quad (22a)$$

$$\frac{\partial \mu}{\partial \xi_m} = \sum_{j=1}^{s}\sum_{i=1}^{m_j} \frac{\partial \mu}{\partial y_i^{(j)}} \frac{\partial y_i^{(j)}}{\partial \xi_m} \tag{22b}$$

and

$$\frac{\partial^2 \mu}{\partial x_l \partial x_m} = \sum_{j'=1}^{s}\sum_{i'=1}^{m_{j'}}\sum_{j=1}^{s}\sum_{i=1}^{m_j} \frac{\partial^2 \mu}{\partial y_{i'}^{(j')} \partial y_i^{(j)}} \frac{\partial y_i^{(j)}}{\partial x_l} \frac{\partial y_{i'}^{(j')}}{\partial x_m} + \sum_{j=1}^{s}\sum_{i=1}^{m_j} \frac{\partial \mu}{\partial y_i^{(j)}} \frac{\partial^2 y_i^{(j)}}{\partial x_l \partial x_m} \tag{23a}$$

$$\frac{\partial^2 \mu}{\partial x_l \partial \xi_m} = \sum_{j'=1}^{s}\sum_{i'=1}^{m_{j'}}\sum_{j=1}^{s}\sum_{i=1}^{m_j} \frac{\partial^2 \mu}{\partial y_{i'}^{(j')} \partial y_i^{(j)}} \frac{\partial y_i^{(j)}}{\partial x_l} \frac{\partial y_{i'}^{(j')}}{\partial \xi_m} + \sum_{j=1}^{s}\sum_{i=1}^{m_j} \frac{\partial \mu}{\partial y_i^{(j)}} \frac{\partial^2 y_i^{(j)}}{\partial x_l \partial \xi_m} \tag{23b}$$

$$\frac{\partial^2 \mu}{\partial \xi_l \partial \xi_m} = \sum_{j'=1}^{s}\sum_{i'=1}^{m_{j'}}\sum_{j=1}^{s}\sum_{i=1}^{m_j} \frac{\partial^2 \mu}{\partial y_{i'}^{(j')} \partial y_i^{(j)}} \frac{\partial y_i^{(j)}}{\partial \xi_l} \frac{\partial y_{i'}^{(j')}}{\partial \xi_m} + \sum_{j=1}^{s}\sum_{i=1}^{m_j} \frac{\partial \mu}{\partial y_i^{(j)}} \frac{\partial^2 y_i^{(j)}}{\partial \xi_l \partial \xi_m} \tag{23c}$$

where the first and second derivatives of $\mu\left(\left\{y_i^{(j)}\right\}\right)$ with respect to site fractions, $\frac{\partial \mu}{\partial y_i^{(j)}}$ and $\frac{\partial^2 \mu}{\partial y_{i'}^{(j')} \partial y_i^{(j)}}$, can again be obtained analytically or numerically from the CALPHAD-based thermodynamic database of the system. The remaining derivatives to be calculated are $\frac{\partial y_i^{(j)}}{\partial x_l}$, $\frac{\partial y_i^{(j)}}{\partial \xi_l}$, $\frac{\partial^2 y_i^{(j)}}{\partial x_l \partial x_m}$, $\frac{\partial^2 y_i^{(j)}}{\partial x_l \partial \xi_m}$ and $\frac{\partial^2 y_i^{(j)}}{\partial \xi_l \partial \xi_m}$, all of which can be calculated in matrix form following $\mathbf{v}=\mathbf{B}^{-1}\mathbf{f}$, where the coefficient matrix $\mathbf{B}$ remains the same as Eq. (18), while the vectors $\mathbf{v}$ and $\mathbf{f}$ are different for different purposes, as listed in Table 2.

Table 2: The specific forms of vectors $\mathbf{v}$ and $\mathbf{f}$ for calculating site fractions, as well as first and second derivatives of site fractions with respect to compositions and order parameters using $\mathbf{v}=\mathbf{B}^{-1}\mathbf{f}$

| Cases | v | f |
|---|---|---|
| $y_i^{(j)}$ | $v_{i+\sum_{p=1}^{j-1} m_p} = y_i^{(j)}$ | See Eq. (19) |
| $\frac{\partial y_i^{(j)}}{\partial x_l}$ | $v_{i+\sum_{p=1}^{j-1} m_p} = \frac{\partial y_i^{(j)}}{\partial x_l}$ | $f_r = \begin{cases} M; \ r-s-1 = l \\ 0; else \end{cases}$ |
| $\dfrac{\partial y_i^{(j)}}{\partial \xi_m}$ | $v_{i+\Sigma_{p=1}^{j-1} m_p} = \dfrac{\partial y_i^{(j)}}{\partial \xi_m}$ | $f_r = \begin{cases} w^m; \ r - c = m, \{R_p^q\}^m \neq null \\ 1; \ r - c = m, \{R_p^q\}^m = null \\ 0; else \end{cases}$ |
| $\dfrac{\partial^2 y_i^{(j)}}{\partial x_l \partial x_m}$ | $v_{i+\Sigma_{p=1}^{j-1} m_p} = \dfrac{\partial^2 y_i^{(j)}}{\partial x_l \partial x_m}$ | $f_r = \begin{cases} \dfrac{\partial M}{\partial x_m}; \ r - s - 1 = l \neq m \\ \dfrac{\partial M}{\partial x_l}; \ r - s - 1 = m \neq l \\ 2\dfrac{\partial M}{\partial x_m}; \ r - s - 1 = l = m \\ 0; else \end{cases}$ |
| $\dfrac{\partial^2 y_i^{(j)}}{\partial x_l \partial \xi_m}$ | $v_{i+\Sigma_{p=1}^{j-1} m_p} = \dfrac{\partial^2 y_i^{(j)}}{\partial x_l \partial \xi_m}$ | $f_r = \begin{cases} \dfrac{\partial M}{\partial \xi_m}; \ r - s - 1 = l \\ 0; else \end{cases}$ |
| $\dfrac{\partial^2 y_i^{(j)}}{\partial \xi_l \partial \xi_m}$ | $v_{i+\Sigma_{p=1}^{j-1} m_p} = \dfrac{\partial^2 y_i^{(j)}}{\partial \xi_l \partial \xi_m}$ | $0$ |

Here $w^m = \sum_{p,q \in \{R_p^q\}^m} y_p^{(q)} + \sum_{p',q' \in \{P_{p'}^{q'}\}^m} y_{p'}^{(q')}$ is the sum of the site fractions of the reacting and product species of the $m$-th internal reaction.

With this strategy, one can numerically calculate the values of site fractions $\{y_i^{(j)}\}$, the chemical potential $\mu$, and derivatives of $\mu$, given the values of compositions $\{x_l\}$ and IPOPs $\{\xi_m\}$, and the analytical expression (or numerical values) of $\mu\left(\{y_i^{(j)}\}\right)$ and its derivatives.

The overall procedure for the variable conversion is shown in the flowchart in Figure 6. Based on the flowchart and the equations shown above, a FORTRAN program has been developed, which is applicable to general nonstoichiometric phases described by the sublattice model.

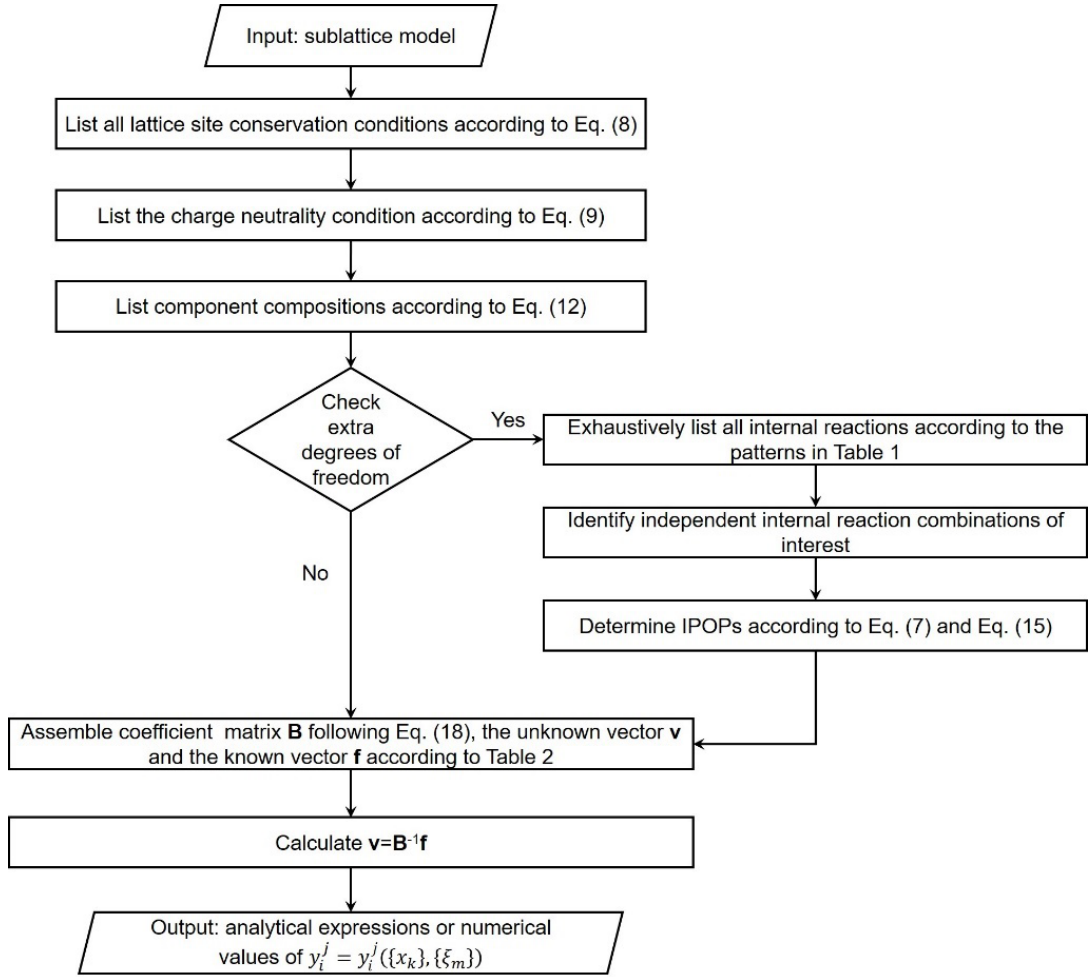

Figure 6: Flowchart for the variable conversion from site fractions to compositions and internal process order parameters.

## 3. Application example

In this section, we show the application of this general conversion strategy to two realistic nonstoichiometric phases, i.e., the Laves phase in the $Al_x$CrNbTiVZr alloy system, and the perovskite phase in the La-Sr-Mn-O system.

### 3.1 Intermetallic Laves phase in high entropy Al-Cr-Nb-Ti-V-Zr system

In nonstoichiometric intermetallic phases, the most common internal processes are atom exchange reactions. For example, in the $Al_x$CrNbTiVZr ($x$ varies from 0.5 to 1.5) high entropy alloy system, C14 Laves phase with a topologically close-packed hexagonal MgZn$_2$ structure has been reported[14]. The C14 Laves phase can be described by the sublattice model (Al, Cr, Nb, Ti, V, Zr)$_2$(Al, Cr, Nb, Ti, V, Zr)$_1$, with 2 sublattices and 12 sublattice sites[15]. There are only 5 independent components in this phase and 5 independent internal atom exchange reactions.

Following the procedures above, we first list all the external constraints. The two lattice site conservation conditions are

$$y_{Al}^{(1)} + y_{Cr}^{(1)} + y_{Nb}^{(1)} + y_{Ti}^{(1)} + y_{V}^{(1)} + y_{Zr}^{(1)} = 1 \tag{24a}$$
$$y_{Al}^{(2)} + y_{Cr}^{(2)} + y_{Nb}^{(2)} + y_{Ti}^{(2)} + y_{V}^{(2)} + y_{Zr}^{(2)} = 1 \tag{24b}$$

Since all the sublattice species (atoms of metallic elements) are charge-neutral species, there is no need to introduce a charge neutrality equation. The Laves phase involves 6 components Al, Cr, Nb, Ti, V, Zr, and they are constrained by $x_{Al} + x_{Cr} + x_{Nb} + x_{Ti} + x_V + x_{Zr} = 1$, so there are only five independent components. We choose Cr, Nb, Ti, V, Zr, and their compositions are defined as

$$x_{Cr} = \frac{2y_{Cr}^{(1)} + y_{Cr}^{(2)}}{M_{Laves}} \tag{25a}$$

$$x_{Nb} = \frac{2y_{Nb}^{(1)} + y_{Nb}^{(2)}}{M_{Laves}} \tag{25b}$$

$$x_{Ti} = \frac{2y_{Ti}^{(1)} + y_{Ti}^{(2)}}{M_{Laves}} \tag{25c}$$

$$x_V = \frac{2y_V^{(1)} + y_V^{(2)}}{M_{Laves}} \tag{25d}$$

$$x_{Zr} = \frac{2y_{Zr}^{(1)} + y_{Zr}^{(2)}}{M_{Laves}} \tag{25e}$$

where $M_{Laves} = 2\left(y_{Al}^{(1)} + y_{Cr}^{(1)} + y_{Nb}^{(1)} + y_{Ti}^{(1)} + y_V^{(1)} + y_{Zr}^{(1)}\right) + \left(y_{Al}^{(2)} + y_{Cr}^{(2)} + y_{Nb}^{(2)} + y_{Ti}^{(2)} + y_V^{(2)} + y_{Zr}^{(2)}\right)$ is the number of moles of species in a formula unit of the Laves phase. According to Eq. (12), these equations can be re-arranged as

$$-2x_{Cr}\left(y_{Al}^{(1)} + y_{Nb}^{(1)} + y_{Ti}^{(1)} + y_V^{(1)} + y_{Zr}^{(1)}\right)$$
$$- x_{Cr}\left(y_{Al}^{(2)} + y_{Nb}^{(2)} + y_{Ti}^{(2)} + y_V^{(2)} + y_{Zr}^{(2)}\right) + 2(1 - x_{Cr})y_{Cr}^{(1)} \quad (26a)$$
$$+ (1 - x_{Cr})y_{Cr}^{(2)} = 0$$

$$-2x_{Nb}\left(y_{Al}^{(1)} + y_{Cr}^{(1)} + y_{Ti}^{(1)} + y_V^{(1)} + y_{Zr}^{(1)}\right)$$
$$- x_{Nb}\left(y_{Al}^{(2)} + y_{Cr}^{(2)} + y_{Ti}^{(2)} + y_V^{(2)} + y_{Zr}^{(2)}\right) + 2(1 - x_{Nb})y_{Nb}^{(1)} \quad (26b)$$
$$+ (1 - x_{Nb})y_{Nb}^{(2)} = 0$$

$$-2x_{Ti}\left(y_{Al}^{(1)} + y_{Nb}^{(1)} + y_{Cr}^{(1)} + y_V^{(1)} + y_{Zr}^{(1)}\right)$$
$$- x_{Ti}\left(y_{Al}^{(2)} + y_{Nb}^{(2)} + y_{Cr}^{(2)} + y_V^{(2)} + y_{Zr}^{(2)}\right) + 2(1 - x_{Ti})y_{Ti}^{(1)} \quad (26c)$$
$$+ (1 - x_{Ti})y_{Ti}^{(2)} = 0$$

$$-2x_V\left(y_{Al}^{(1)} + y_{Nb}^{(1)} + y_{Ti}^{(1)} + y_{Cr}^{(1)} + y_{Zr}^{(1)}\right)$$
$$- x_V\left(y_{Al}^{(2)} + y_{Nb}^{(2)} + y_{Ti}^{(2)} + y_{Cr}^{(2)} + y_{Zr}^{(2)}\right) + 2(1 - x_V)y_V^{(1)} \quad (26d)$$
$$+ (1 - x_V)y_V^{(2)} = 0$$

$$-2x_{Zr}\left(y_{Al}^{(1)} + y_{Nb}^{(1)} + y_{Ti}^{(1)} + y_V^{(1)} + y_{Cr}^{(1)}\right)$$
$$- x_{Zr}\left(y_{Al}^{(2)} + y_{Nb}^{(2)} + y_{Ti}^{(2)} + y_V^{(2)} + y_{Cr}^{(2)}\right) + 2(1 - x_{Zr})y_{Zr}^{(1)} \quad (26e)$$
$$+ (1 - x_{Zr})y_{Zr}^{(2)} = 0$$

In addition to the external constraints, the Laves involves multiple internal processes. By matching with the "patterns" in Table 1, we find that the atoms of all the 6 components are present in both sublattices, so there are $C_6^2=15$ species pairs to allow for the atom exchange reactions. No other internal processes can be identified. These 15 species pairs are (Al, Cr), (Al, Nb), (Al, Ti), (Al, V), (Al, Zr), (Cr, Nb), (Cr, Ti), (Cr, V), (Cr, Zr), (Nb, Ti), (Nb, V), (Nb, Zr), (Ti, V), (Ti, Zr) and (V, Zr). Similar to Eq. (11), we can list all these atom exchange reactions in matrix form and determine that the rank of the coefficient matrix **A** is 5. There are $C_{15}^5=3003$ possible ways to select 5 out of the 15 reactions, while only 1296 of them can give 5 independent reactions, which can be identified and listed using the developed Fortran program. For simplicity, we only consider one of the 1296 independent combinations, i.e.,

$$Al^{(1)} + Cr^{(2)} = Al^{(2)} + Cr^{(1)} \quad (27a)$$
$$Al^{(1)} + Nb^{(2)} = Al^{(2)} + Nb^{(1)} \quad (27b)$$

$$Al^{(1)} + Ti^{(2)} = Al^{(2)} + Ti^{(1)} \tag{27c}$$
$$Al^{(1)} + V^{(2)} = Al^{(2)} + V^{(1)} \tag{27d}$$
$$Al^{(1)} + Zr^{(2)} = Al^{(2)} + Zr^{(1)} \tag{27e}$$

The corresponding IPOPs can then be defined as:

$$\xi_1 = \frac{y_{Al}^{(2)} + y_{Cr}^{(1)}}{y_{Al}^{(2)} + y_{Cr}^{(1)} + y_{Al}^{(1)} + y_{Cr}^{(2)}} \tag{28a}$$

$$\xi_2 = \frac{y_{Al}^{(2)} + y_{Nb}^{(1)}}{y_{Al}^{(2)} + y_{Nb}^{(1)} + y_{Al}^{(1)} + y_{Nb}^{(2)}} \tag{28b}$$

$$\xi_3 = \frac{y_{Al}^{(2)} + y_{Ti}^{(1)}}{y_{Al}^{(2)} + y_{Ti}^{(1)} + y_{Al}^{(1)} + y_{Ti}^{(2)}} \tag{28c}$$

$$\xi_4 = \frac{y_{Al}^{(2)} + y_{V}^{(1)}}{y_{Al}^{(2)} + y_{V}^{(1)} + y_{Al}^{(1)} + y_{V}^{(2)}} \tag{28d}$$

$$\xi_5 = \frac{y_{Al}^{(2)} + y_{Zr}^{(1)}}{y_{Al}^{(2)} + y_{Zr}^{(1)} + y_{Al}^{(1)} + y_{Zr}^{(2)}} \tag{28e}$$

which can be re-arranged as

$$(1-\xi_1)y_{Al}^{(2)} + (1-\xi_1)y_{Cr}^{(1)} - \xi_1 y_{Al}^{(1)} - \xi_1 y_{Cr}^{(2)} = 0 \tag{29a}$$
$$(1-\xi_2)y_{Al}^{(2)} + (1-\xi_2)y_{Nb}^{(1)} - \xi_2 y_{Al}^{(1)} - \xi_2 y_{Nb}^{(2)} = 0 \tag{29b}$$
$$(1-\xi_3)y_{Al}^{(2)} + (1-\xi_3)y_{Ti}^{(1)} - \xi_3 y_{Al}^{(1)} - \xi_3 y_{Ti}^{(2)} = 0 \tag{29c}$$
$$(1-\xi_4)y_{Al}^{(2)} + (1-\xi_4)y_{V}^{(1)} - \xi_4 y_{Al}^{(1)} - \xi_4 y_{V}^{(2)} = 0 \tag{29d}$$
$$(1-\xi_5)y_{Al}^{(2)} + (1-\xi_5)y_{Zr}^{(1)} - \xi_5 y_{Al}^{(1)} - \xi_5 y_{Zr}^{(2)} = 0 \tag{29e}$$

We can then collect Eqs. (24), (26) and (29) into the matrix form $\mathbf{B}^{Laves}\mathbf{v}^{Laves}=\mathbf{f}^{Laves}$ as

$$\begin{pmatrix} 1 & 1 & 1 & 1 & 1 & 1 & 0 & 0 & 0 & 0 & 0 & 0 \\ 0 & 0 & 0 & 0 & 0 & 0 & 1 & 1 & 1 & 1 & 1 & 1 \\ -2x_{Cr} & 2(1-x_{Cr}) & -2x_{Cr} & -2x_{Cr} & -2x_{Cr} & -2x_{Cr} & -x_{Cr} & (1-x_{Cr}) & -x_{Cr} & -x_{Cr} & -x_{Cr} & -x_{Cr} \\ -2x_{Nb} & -2x_{Nb} & 2(1-x_{Nb}) & -2x_{Nb} & -2x_{Nb} & -2x_{Nb} & -x_{Nb} & -x_{Nb} & (1-x_{Nb}) & -x_{Nb} & -x_{Nb} & -x_{Nb} \\ -2x_{Ti} & -2x_{Ti} & -2x_{Ti} & 2(1-x_{Ti}) & -2x_{Ti} & -2x_{Ti} & -x_{Ti} & -x_{Ti} & -x_{Ti} & (1-x_{Ti}) & -x_{Ti} & -x_{Ti} \\ -2x_{V} & -2x_{V} & -2x_{V} & -2x_{V} & 2(1-x_{V}) & -2x_{V} & -x_{V} & -x_{V} & -x_{V} & -x_{V} & (1-x_{V}) & -x_{V} \\ -2x_{Zr} & -2x_{Zr} & -2x_{Zr} & -2x_{Zr} & -2x_{Zr} & 2(1-x_{Zr}) & -x_{Zr} & -x_{Zr} & -x_{Zr} & -x_{Zr} & -x_{Zr} & (1-x_{Zr}) \\ -\xi_1 & 1-\xi_1 & 0 & 0 & 0 & 0 & 1-\xi_1 & -\xi_1 & 0 & 0 & 0 & 0 \\ -\xi_2 & 0 & 1-\xi_2 & 0 & 0 & 0 & 1-\xi_2 & 0 & -\xi_2 & 0 & 0 & 0 \\ -\xi_3 & 0 & 0 & 1-\xi_3 & 0 & 0 & 1-\xi_3 & 0 & 0 & -\xi_3 & 0 & 0 \\ -\xi_4 & 0 & 0 & 0 & 1-\xi_4 & 0 & 1-\xi_4 & 0 & 0 & 0 & -\xi_4 & 0 \\ -\xi_5 & 0 & 0 & 0 & 0 & 1-\xi_5 & 1-\xi_5 & 0 & 0 & 0 & 0 & -\xi_5 \end{pmatrix} \begin{pmatrix} y_{Al}^1 \\ y_{Cr}^1 \\ y_{Nb}^1 \\ y_{Ti}^1 \\ y_{V}^1 \\ y_{Zr}^1 \\ y_{Al}^2 \\ y_{Cr}^2 \\ y_{Nb}^2 \\ y_{Ti}^2 \\ y_{V}^2 \\ y_{Zr}^2 \end{pmatrix} = \begin{pmatrix} 1 \\ 1 \\ 0 \\ 0 \\ 0 \\ 0 \\ 0 \\ 0 \\ 0 \\ 0 \\ 0 \\ 0 \end{pmatrix} \tag{30}$$

Then the site fractions $\mathbf{v}^{Laves} = \left(y_{Al}^{(1)}, y_{Cr}^{(1)}, y_{Nb}^{(1)}, y_{Ti}^{(1)}, y_{V}^{(1)}, y_{Zr}^{(1)}, y_{Al}^{(2)}, y_{Cr}^{(2)}, y_{Nb}^{(2)}, y_{Ti}^{(2)}, y_{V}^{(2)}, y_{Zr}^{(2)}\right)^T$ can be obtained by $\mathbf{v}^{Laves} = (\mathbf{B}^{Laves})^{-1}\mathbf{f}^{Laves}$. The detailed analytical expressions for the site fractions are listed in Supplementary Materials S1.

### 3.2 Perovskite phase in La-Sr-Mn-O system

We then apply the proposed procedure and numerical tools to a nonstoichiometric oxide, i.e., the $La_xSr_{1-x}MnO_3$ (LSM) perovskite phase, which involves Mn cations with different valences and possible internal redox reactions. LSM is a common cathode material for solid oxide fuel cells (SOFCs). This nonstoichiometric phase is described by the sublattice model $(La^{3+},Sr^{2+},Mn^{3+},Va)_1(Mn^{2+},Mn^{3+},Mn^{4+},Va)_1(O^{2-},Va)_3$ with 3 sublattices and 10 site fractions[16]. Following the procedures above, we first list the external constraints, including 3 lattice conservation conditions,

$$y^{(1)}_{La^{3+}} + y^{(1)}_{Sr^{2+}} + y^{(1)}_{Mn^{3+}} + y^{(1)}_{Va} = 1 \tag{31a}$$
$$y^{(2)}_{Mn^{2+}} + y^{(2)}_{Mn^{3+}} + y^{(2)}_{Mn^{4+}} + y^{(2)}_{Va} = 1 \tag{31b}$$
$$y^{(3)}_{O^{2-}} + y^{(3)}_{Va} = 1 \tag{31c}$$

one charge neutrality condition,

$$3y^{(1)}_{La^{3+}} + 2y^{(1)}_{Sr^{2+}} + 3y^{(1)}_{Mn^{3+}} + 2y^{(2)}_{Mn^{2+}} + 3y^{(2)}_{Mn^{3+}} + 4y^{(2)}_{Mn^{4+}} - 6y^{(3)}_{O^{2-}} = 0 \tag{32}$$

four relations among compositions and site fractions,

$$x_{La} = \frac{y^{(1)}_{La^{3+}}}{M_{LSM}} \tag{33a}$$

$$x_{Sr} = \frac{y^{(1)}_{Sr^{2+}}}{M_{LSM}} \tag{33b}$$

$$x_{Mn} = \frac{y^{(1)}_{Mn^{3+}} + y^{(2)}_{Mn^{2+}} + y^{(2)}_{Mn^{3+}} + y^{(2)}_{Mn^{4+}}}{M_{LSM}} \tag{33c}$$

$$x_O = \frac{3y^{(3)}_{O^{2-}}}{M_{LSM}} \tag{33d}$$

here $M_{LSM} = y^{(1)}_{La^{3+}} + y^{(1)}_{Sr^{2+}} + y^{(1)}_{Mn^{3+}} + y^{(2)}_{Mn^{2+}} + y^{(2)}_{Mn^{3+}} + y^{(2)}_{Mn^{4+}} + 3y^{(3)}_{O^{2-}}$ is the number of moles of species in a formula unit of LSM. Due to the mass conservation condition $x_{La} + x_{Sr} + x_{Mn} + x_O = 1$, there are at most 3 independent compositions. Note that in the current example, Mn is an independent component, while $Mn^{2+}$, $Mn^{3+}$ and $Mn^{4+}$ are all species of Mn and should not be treated as three independent components.

We then deal with the internal constraints. To start with, we list all the internal reactions following the "patterns" in Table 1. These internal reactions are:

$$(Mn^{3+})^{(1)} + (Va)^{(2)} \rightarrow (Mn^{3+})^{(2)} + (Va)^{(1)} \quad (34a)$$
$$(Mn^{2+})^{(2)} + (Mn^{4+})^{(2)} \rightarrow 2(Mn^{3+})^{(2)} \quad (34b)$$
$$null \rightarrow Va^{(1)} + Va^{(2)} + 3Va^{(3)} \quad (34c)$$

These reactions can be written in matrix form **Au=0** as:

$$\begin{pmatrix} 1 & -1 & 0 & -1 & 0 & 1 & 0 \\ 0 & 0 & 1 & -2 & 1 & 0 & 0 \\ 0 & -1 & 0 & 0 & 0 & -1 & -3 \end{pmatrix} \begin{pmatrix} (Mn^{3+})^{(1)} \\ Va^{(1)} \\ (Mn^{2+})^{(2)} \\ (Mn^{3+})^{(2)} \\ (Mn^{4+})^{(2)} \\ Va^{(2)} \\ Va^{(3)} \end{pmatrix} = \mathbf{0} \quad (35)$$

where the rank of the 3×7 coefficient matrix **A** is 3, indicating the three internal reactions are independent of each other. Therefore, the extra degrees of freedom can be fixed: 10 site fractions = 3 lattice conservation conditions + 1 charge neutrality condition + 3 relations with independent compositions (e.g., La, Sr, Mn) + 3 internal reactions. The corresponding IPOPs can be defined as

$$\xi_1 = \frac{y^{(2)}_{Mn^{3+}} + y^{(1)}_{Va}}{y^{(1)}_{Mn^{3+}} + y^{(2)}_{Va} + y^{(2)}_{Mn^{3+}} + y^{(1)}_{Va}} \quad (36a)$$

$$\xi_2 = \frac{y^{(2)}_{Mn^{3+}}}{y^{(2)}_{Mn^{2+}} + y^{(2)}_{Mn^{4+}} + y^{(2)}_{Mn^{3+}}} \quad (36b)$$

$$\xi_3 = \frac{1}{5}\left(y_{Va}^{(1)} + y_{Va}^{(2)} + 3y_{Va}^{(3)}\right) \tag{36c}$$

We can then express the 10×10 coefficient matrix **B** according to Eq. (18):

$$\mathbf{B}^{LSM} = \begin{pmatrix}
1 & 1 & 1 & & 1 & 0 & 0 & & 0 & 0 & 0 & 0 \\
0 & 0 & 0 & & 0 & 1 & 1 & & 1 & 1 & 0 & 0 \\
0 & 0 & 0 & & 0 & 0 & 0 & & 0 & 0 & 1 & 1 \\
3 & 2 & 3 & 0 & 2 & 3 & & 4 & 0 & -6 & 0 \\
1-x_{La} & -x_{La} & -x_{La} & 0 & -x_{La} & -x_{La} & & -x_{La} & 0 & -3x_{La} & 0 \\
-x_{Sr} & 1-x_{Sr} & -x_{Sr} & 0 & -x_{Sr} & -x_{Sr} & & -x_{Sr} & 0 & -3x_{Sr} & 0 \\
-x_{Mn} & -x_{Mn} & 1-x_{Mn} & 0 & 1-x_{Mn} & 1-x_{Mn} & 1-x_{Mn} & 0 & -3x_{Mn} & 0 \\
0 & 0 & -\xi_1 & 1-\xi_1 & 0 & 1-\xi_1 & 0 & -\xi_1 & 0 & 0 \\
0 & 0 & 0 & 0 & -\xi_2 & 1-\xi_2 & -\xi_2 & 0 & 0 & 0 \\
0 & 0 & 0 & \frac{1}{5} & 0 & 0 & 0 & \frac{1}{5} & 0 & \frac{3}{5}
\end{pmatrix} \tag{37}$$

where the first three rows are from the three external lattice site conservation conditions (Eq. (31)), the 4$^{th}$ row is due to the charge neutrality condition (Eq. (32)), the 5$^{th}$~7$^{th}$ rows are from the definition of the three independent compositions (Eq. (33)), while the 8$^{th}$~10$^{th}$ rows are from the definition of IPOPs (Eq. (36)). The species in the unknown vector $\mathbf{y}^{LSM}$ follow their orders in the sublattice model, i.e., $\mathbf{y}^{LSM} = \left(y_{La^{3+}}^{(1)}, y_{Sr^{2+}}^{(1)}, y_{Mn^{3+}}^{(1)}, y_{Va}^{(1)}, y_{Mn^{2+}}^{(2)}, y_{Mn^{3+}}^{(2)}, y_{Mn^{4+}}^{(2)}, y_{Va}^{(2)}, y_{O^{2-}}^{(3)}, y_{Va}^{(3)}\right)^T$. The known vector $\mathbf{f}^{LSM}$ for converting site fractions can be determined according to Eq. (19), which is $\mathbf{f}^{LSM} = (1,1,1,0,0,0,0,0,0,\xi_3)^T$. Other vectors for calculating the first and second derivatives of site fractions with respect to compositions and order parameters can also be calculated according to Table 2.

With these expressions, one can obtain the analytical expressions of site fractions and therefore $\mu^{LSM}$ and its first and second derivatives as a function of $\{x_l\}$ and $\{\xi_m\}$ by taking $\mathbf{y}^{LSM} = (\mathbf{B}^{LSM})^{-1}\mathbf{f}^{LSM}$ and plugging the expressions of $\mathbf{y}^{LSM}$ into $\mu^{LSM}(\{y_i^j\})$. As such, we can convert $\mu^{LSM}\left(y_{La^{3+}}^{(1)}, y_{Sr^{2+}}^{(1)}, y_{Mn^{3+}}^{(1)}, y_{Va}^{(1)}, y_{Mn^{2+}}^{(2)}, y_{Mn^{3+}}^{(2)}, y_{Mn^{4+}}^{(2)}, y_{Va}^{(2)}, y_{O^{2-}}^{(3)}, y_{Va}^{(3)}\right)$ to $\mu^{LSM}(x_{La}, x_{Sr}, x_{Mn}, \xi_1, \xi_2, \xi_3)$. The detailed $\mathbf{y}^{LSM}$ expressions are shown in Supplementary

Materials S2. As an alternative, one can obtain numerical values of site fractions, and its first and second derivatives using the general equation $\mathbf{y}=\mathbf{B}^{-1}\mathbf{f}$ under any given $\{x_l\}$ ($l$=La,Sr,Mn) and $\{\xi_m\}$ ($m$=1,2,3) values. For example, let us consider the LSM phase with a composition of $(La_{0.8}Sr_{0.2})_{0.95}MnO_3$, i.e., $x_{La} = 0.153535$, $x_{Sr} = 0.038384$, $x_{Mn} = 0.20202$, which is a typical composition of LSM in SOFC cathode applications. At $T$=1600 K and $p$=1 atm, using the available thermodynamic database[16], we can find the site fractions under internal equilibrium using OpenCalphad, as listed in Table 3. Based on the definitions in Eq. (36), we can obtain the IPOP values, as listed in Table 4.

Table 3: Site fractions of the LSM phase with a composition of $(La_{0.8}Sr_{0.2})_{0.95}MnO_3$ under internal equilibrium at $T$=1600 K and $p$=1 atm.

| Variables | Values |
|---|---|
| $y^{(1)}_{La^{3+}}$ | 0.759996 |
| $y^{(1)}_{Sr^{2+}}$ | 0.189999 |
| $y^{(1)}_{Mn^{3+}}$ | $2.67838 \times 10^{-3}$ |
| $y^{(1)}_{Va}$ | 0.0473266 |
| $y^{(2)}_{Mn^{2+}}$ | 0.0185997 |
| $y^{(2)}_{Mn^{3+}}$ | 0.620105 |
| $y^{(2)}_{Mn^{4+}}$ | 0.358613 |
| $y^{(2)}_{Va}$ | $2.6826 \times 10^{-3}$ |
| $y^{(3)}_{O^{2-}}$ | 0.999998 |
| $y^{(3)}_{Va}$ | $2.23711 \times 10^{-6}$ |

Table 4: The IPOP values for the site fractions in Table 3, and their physical meanings.

| Variables | Values | Physical meaning |
|---|---|---|
| $\xi_1$ | 0.992032 | Extent of Reaction (34a) |
| $\xi_2$ | 0.621773 | Extent of Reaction (34b) |
| $\xi_3$ | 0.0100032 | Vacancy concentration, or extent of Reaction (34c) |

We can further verify the internal equilibrium by testing if Eq. (2), i.e.,

$$D_j = -\left(\frac{\partial\left(\frac{\mu^{LSM}}{M_{LSM}}\right)}{\partial \xi_j}\right)_{T,p,x_i,\xi_{k\neq j}} = -\frac{M_{LSM}\left(\frac{\partial \mu^{LSM}}{\partial \xi_j}\right)_{T,p,x_i,\xi_{k\neq j}} - \mu^{LSM}\frac{\partial M_{LSM}}{\partial \xi_j}}{M_{LSM}^2} = 0 \text{ are all satisfied for}$$

$j$=1,2,3 after the variable conversion. To do so, according to Eq. (22), we first obtain all the $\frac{\partial \mu^{LSM}}{\partial y_i^{(j)}}$ (for all $i$s and $j$s) values using OpenCalphad as shown in Table 5. Then we need to calculate $\frac{\partial y_i^{(j)}}{\partial \xi_m}$ values (for all $i$s, $j$s and $m$s) according to the 3$^{rd}$ case (4$^{th}$ row) in Table 2, which again can be obtained by **y**=**B**$^{-1}$**f**, with **y** being a 10×3 matrix (10 site fractions, 3 IPOPs) of the unknown $\frac{\partial y_i^{(j)}}{\partial \xi_m}$ values, **B** being the known matrix in Eq. (37) under given $\{x_l\}$ ($l$=La,Sr,Mn) and $\{\xi_m\}$ ($m$=1,2,3) values, and **f** being a known 10×3 matrix:

$$\mathbf{f} = \begin{pmatrix} 0 & 0 & 0 & 0 & 0 & 0 & 0 & y_{Mn^{3+}}^{(1)} + y_{Va}^{(1)} + y_{Mn^{3+}}^{(2)} + y_{Va}^{(3)} & 0 & 0 \\ 0 & 0 & 0 & 0 & 0 & 0 & 0 & 0 & y_{Mn^{2+}}^{(2)} + y_{Mn^{3+}}^{(2)} + y_{Mn^{4+}}^{(2)} & 0 \\ 0 & 0 & 0 & 0 & 0 & 0 & 0 & 0 & 0 & 1 \end{pmatrix}^T \quad (38)$$

The calculated $\frac{\partial y_i^{(j)}}{\partial \xi_m}$ values are listed in Table 6.

Table 5: $\mu$ and $\frac{\partial \mu}{\partial y_i^{(j)}}$ values of the (La$_{0.8}$Sr$_{0.2}$)$_{0.95}$MnO$_3$ perovskite under internal equilibrium at $T$=1600 K and $p$=1 atm, obtained using OpenCalphad with the database in [16].

| Variables | Values (J/mol) | Variables | Values (J/mol) | Variables | Values (J/mol) |
|---|---|---|---|---|---|
| $\frac{\partial \mu^{LSM}}{\partial y_{La^{3+}}^{(1)}}$ | -1.82006211×10$^6$ | $\frac{\partial \mu^{LSM}}{\partial y_{Mn^{2+}}^{(2)}}$ | -1.62383287×10$^6$ | $\frac{\partial \mu^{LSM}}{\partial y_{O^{2-}}^{(3)}}$ | -1.67126901×10$^6$ |
| $\frac{\partial \mu^{LSM}}{\partial y_{Sr^{2+}}^{(1)}}$ | -1.55176564×10$^6$ | $\frac{\partial \mu^{LSM}}{\partial y_{Mn^{3+}}^{(2)}}$ | -1.68786791×10$^6$ | $\frac{\partial \mu^{LSM}}{\partial y_{Va}^{(3)}}$ | -1.63214658×10$^6$ |
| $\frac{\partial \mu^{LSM}}{\partial y_{Mn^{3+}}^{(1)}}$ | -1.33653617×10$^6$ | $\frac{\partial \mu^{LSM}}{\partial y_{Mn^{3+}}^{(2)}}$ | -1.75190295×10$^6$ | $\mu^{LSM}$ | -1.73031445×10$^6$ |
| $\frac{\partial \mu^{LSM}}{\partial y_{Va}^{(1)}}$ | -6.89841721×10$^5$ | $\frac{\partial \mu^{LSM}}{\partial y_{Va}^{(2)}}$ | -1.04117346×10$^6$ | | |

Table 6: $\frac{\partial y_i^{(j)}}{\partial \xi_m}$ values of the $(La_{0.8}Sr_{0.2})_{0.95}MnO_3$ perovskite under internal equilibrium at $T$=1600 K and $p$=1 atm, obtained using Table 2.

| Variables | Values | Variables | Values | Variables | Values |
|---|---|---|---|---|---|
| $\frac{\partial y_{La^{3+}}^{(1)}}{\partial \xi_1}$ | 0 | $\frac{\partial y_{La^{3+}}^{(1)}}{\partial \xi_2}$ | 0 | $\frac{\partial y_{La^{3+}}^{(1)}}{\partial \xi_3}$ | -0.767675 |
| $\frac{\partial y_{Sr^{2+}}^{(1)}}{\partial \xi_1}$ | 0 | $\frac{\partial y_{Sr^{2+}}^{(1)}}{\partial \xi_2}$ | 0 | $\frac{\partial y_{Sr^{2+}}^{(1)}}{\partial \xi_3}$ | -0.191919 |
| $\frac{\partial y_{Mn^{3+}}^{(1)}}{\partial \xi_1}$ | -0.336904 | $\frac{\partial y_{Mn^{3+}}^{(1)}}{\partial \xi_2}$ | 0.00397931 | $\frac{\partial y_{Mn^{3+}}^{(1)}}{\partial \xi_3}$ | -0.500459 |
| $\frac{\partial y_{Va}^{(1)}}{\partial \xi_1}$ | 0.336904 | $\frac{\partial y_{Va}^{(1)}}{\partial \xi_2}$ | -0.00397931 | $\frac{\partial y_{Va}^{(1)}}{\partial \xi_3}$ | 1.46005 |
| $\frac{\partial y_{Mn^{2+}}^{(2)}}{\partial \xi_1}$ | 0.0637131 | $\frac{\partial y_{Mn^{2+}}^{(2)}}{\partial \xi_2}$ | -0.499411 | $\frac{\partial y_{Mn^{2+}}^{(2)}}{\partial \xi_3}$ | 0.0753445 |
| $\frac{\partial y_{Mn^{3+}}^{(2)}}{\partial \xi_1}$ | 0.209478 | $\frac{\partial y_{Mn^{3+}}^{(2)}}{\partial \xi_2}$ | 0.994843 | $\frac{\partial y_{Mn^{3+}}^{(2)}}{\partial \xi_3}$ | -0.316881 |
| $\frac{\partial y_{Mn^{4+}}^{(2)}}{\partial \xi_1}$ | 0.0637131 | $\frac{\partial y_{Mn^{4+}}^{(2)}}{\partial \xi_2}$ | -0.499411 | $\frac{\partial y_{Mn^{4+}}^{(2)}}{\partial \xi_3}$ | -0.268104 |
| $\frac{\partial y_{Va}^{(2)}}{\partial \xi_1}$ | -0.336904 | $\frac{\partial y_{Va}^{(2)}}{\partial \xi_2}$ | 0.00397931 | $\frac{\partial y_{Va}^{(2)}}{\partial \xi_3}$ | 0.509641 |
| $\frac{\partial y_{O^{2-}}^{(3)}}{\partial \xi_1}$ | 0 | $\frac{\partial y_{O^{2-}}^{(3)}}{\partial \xi_2}$ | 0 | $\frac{\partial y_{O^{2-}}^{(3)}}{\partial \xi_3}$ | -1.0101 |
| $\frac{\partial y_{Va}^{(3)}}{\partial \xi_1}$ | 0 | $\frac{\partial y_{Va}^{(3)}}{\partial \xi_2}$ | 0 | $\frac{\partial y_{Va}^{(3)}}{\partial \xi_3}$ | 1.0101 |

With both $\frac{\partial \mu^{LSM}}{\partial y_i^{(j)}}$ (for all $i$s and $j$s) and $\frac{\partial y_i^{(j)}}{\partial \xi_m}$ (for all $i$s, $j$s and $m$s) values, according to Eq. (22), we can obtain $\left(\frac{\partial \mu^{LSM}}{\partial \xi_j}\right)_{T,p,x_i,\xi_{k \neq j}}$ values. Meanwhile, since $M_{LSM} = y_{La^{3+}}^{(1)} + y_{Sr^{2+}}^{(1)} + y_{Mn^{3+}}^{(1)} + y_{Mn^{2+}}^{(2)} + y_{Mn^{3+}}^{(2)} + y_{Mn^{4+}}^{(2)} + 3y_{O^{2-}}^{(3)} = 5 - y_{Va}^{(1)} - y_{Va}^{(2)} - 3y_{Va}^{(3)}$, the values of $\frac{\partial M_{LSM}}{\partial \xi_j} = -\frac{\partial y_{Va}^{(1)}}{\partial \xi_j} - \frac{\partial y_{Va}^{(2)}}{\partial \xi_j} - 3\frac{\partial y_{Va}^{(3)}}{\partial \xi_j}$ can be obtained using Table 6. We can then obtain

$$D_1 = -\left(\frac{\partial \left(\frac{\mu^{LSM}}{M_{LSM}}\right)}{\partial \xi_1}\right)_{T,p,x_i,\xi_{k \neq 1}} = 0.0683 \, J/mol, \quad D_2 = -\left(\frac{\partial \left(\frac{\mu^{LSM}}{M_{LSM}}\right)}{\partial \xi_1}\right)_{T,p,x_i,\xi_{k \neq 2}} = 0.1057 \, J/mol$$

and $D_3 = -\left(\frac{\partial \left(\frac{\mu^{LSM}}{M_{LSM}}\right)}{\partial \xi_3}\right)_{T,p,x_i,\xi_{k \neq 3}} = 0.861 \, J/mol$. All of these driving forces are sufficiently close to zero compared to the chemical potential and its derivative values (~$10^6$

J/mol) in Table 5. Therefore, within the acceptable numerical error range, the internal equilibrium condition is justified. This verifies our variable conversion approaches in Table 2.

We would like to emphasize that our approach can also be applied to nonequilibrium conditions where the internal equilibrium has not been established. Under such conditions, we need to know the extent of the internal processes in order to perform the variable conversions and related calculations. For example, let us consider the same LSM phase $(La_{0.8}Sr_{0.2})_{0.95}MnO_3$, which has reached internal equilibrium at $T=1600$ K, and is then quenched to $T=1000$ K where the internal processes in Eq. (34) have not proceeded further. The LSM phase is now at a nonequilibrium state with the same site fractions as in Table 3 and the same IPOPs as in Table 4. Following similar procedures above, we can calculate the driving forces of the internal processes as $D_1 = -\left(\frac{\partial \left(\frac{\mu^{LSM}}{M_{LSM}}\right)}{\partial \xi_1}\right)_{T,p,x_i,\xi_{k \neq 1}} = 684.775\ J/mol$, $D_2 = -\left(\frac{\partial \left(\frac{\mu^{LSM}}{M_{LSM}}\right)}{\partial \xi_1}\right)_{T,p,x_i,\xi_{k \neq 2}} = 1005.32\ J/mol$ and $D_3 = -\left(\frac{\partial \left(\frac{\mu^{LSM}}{M_{LSM}}\right)}{\partial \xi_3}\right)_{T,p,x_i,\xi_{k \neq 3}} = -28329\ J/mol$, indicating that reactions (34a) and (34b) tend to proceed further, generating more $Mn^{3+}$ species in the second sublattice, while the reaction (34c) tend to reverse its direction, decreasing the overall vacancy concentration.

### 4. Internal process order parameter database

The current variable conversion procedure and numerical tools are generally applicable to any nonstoichiometric phases described by sublattice models. With such tools, thermodynamic databases for internal processes and internal process order parameters in nonstoichiometric phases can be developed, which can greatly facilitate the variable conversion, thermodynamic analysis and kinetic modeling for these phases. To do so, one can collect all the nonstoichiometric phases described by sublattice models from existing CALPHAD-type thermodynamic databases and identify the phases with internal processes. By applying the tools developed in Section 2, the specific internal processes, the corresponding internal process order parameters, and the specific form of the matrix **B** and vector **f** for variable conversion, can be obtained. For example, Table 7 in the supplementary materials summarizes the findings for all the nonstoichiometric phases with internal processes in the TCAL8 database[17], which can be considered as a sample IPOP database of nonstoichiometric phases in Al-alloys. Note that for nonstoichiometric phases involving multiple components and multiple internal processes, the choices of independent compositions and independent internal processes may not be unique, and we only provide one such choice for each phase in the database. These different choices of independent internal processes are thermodynamically equivalent, but their kinetics (rate constant and activation energy) could be quite different, which we will further explore in our forthcoming studies.

## 5. Summary and perspectives

We developed a general procedure and a set of mathematical tools to convert the sublattice site fractions of a nonstoichiometric phase to a combination of independent component compositions and internal process order parameters. By applying these tools to realistic nonstoichiometric phases from existing thermodynamic databases and literature, we constructed a sample database of internal process order parameters in nonstoichiometric phases, which would facilitate the thermodynamic analysis and kinetic modeling of such phases.

With the identification of internal processes, this study can motivate the development of more accurate thermodynamic models by explicitly considering the constraints and driving forces of the internal processes for nonstoichiometric compounds and materials with redox and defect generation and reaction such as high entropy oxides and radiated materials. Moreover, a nonstoichiometric phase may involve multiple internal processes with different kinetic rates, which may play different roles in the kinetic pathways of potential phase transformations. This could also motivate the investigation of the kinetics of the internal processes and the development of more comprehensive kinetic databases of nonstoichiometric phases.

**Acknowledgement**

Yanzhou Ji and Long-Qing Chen acknowledge the financial support from the Hamer Foundation through the Hamer Professorship at Penn State, as well as the partial site contract support from the United States Department of Energy, National Energy Technology Laboratory to support the US Department of Energy's Fossil Energy Solid Oxide Fuel Cell Program. Yanzhou Ji also acknowledge the start-up funding from The Ohio State University. Yueze Tan acknowledges the support from the National Science Foundation under grant DMR-2011839 through the MRSEC Center for Nanoscale Science of The Pennsylvania State University. The calculations were performed on the Roar supercomputer at Pennsylvania State University and the Ohio Supercomputer Center (OSC).

**Data availability**

Data have been included in supplementary materials. Codes will be made available on request.